\documentclass[aps,prb,twocolumn,superscriptaddress,showpacs,floatfix]{revtex4-1}
\usepackage{graphicx}
\usepackage{amsmath}
\usepackage{amssymb}
\usepackage{url}
\usepackage{float}
\usepackage{color}
\usepackage{natbib}

\begin{document}
\bibliographystyle{apsrev}

\title{Andreev reflection in edge states of time reversal invariant Landau
levels}

\author{K. G. S. H. Gunawardana}
\altaffiliation{Currently at Ames Laboratory, Iowa State University, USA.}
\author{Bruno Uchoa}

\affiliation{Department of Physics and Astronomy, University of Oklahoma, Norman,
OK 73069, USA}

\email{harshakgs@gmail.com, uchoa@ou.edu}

%\selectlanguage{english}%

\date{\today}
\begin{abstract}
We describe the conductance of a normal-superconducting junction in
systems with Landau levels that preserve time reversal symmetry. Those
Landau levels have been observed in strained honeycomb lattices. The
current is carried along the edges in both the normal and superconducting
regions. When the Landau levels in the normal region are half-filled,
Andreev reflection is maximal and the conductance plateaus have a
peak as a function of filling factor. The height of those peaks is
quantized at $4e^{2}/h$. The interface of the junction has Andreev
edge states, which form a coherent superposition of electrons and
holes that can carry a net valley current. We identify
unique experimental signatures for superconductivity in time reversal invariant
Landau levels. 
\end{abstract}


\maketitle
\emph{Introduction. }At zero temperature, Cooper pairs are protected
against phase decoherence by time reversal symmetry (TRS). The most
promising scenario to study the coexistence of superconductivity and
quantum Hall states \cite{Tesanovic} requires a system that creates
well separated Landau levels (LLs) in the complete absence of magnetic
flux \cite{Uchoa}. In strained honeycomb lattices, uniform strain
fields can couple to the electrons as a pseudo-magnetic field oriented
in opposite directions in the two valleys \cite{Guinea,Low}. As observed
experimentally in graphene \cite{Levy,Gomes,Yeh} and in deformed
honeycomb optical lattices \cite{Rechtsman}, this field can produce
sharp LL quantization, and at the same time, zero net magnetic flux
at every lattice site. 

In the usual semi-classical picture, electrons moving in cyclotronic
orbits are reflected as holes at the interface of a superconductor.
At low energy, the holes have opposite
momentum (valley) and the same velocity of the incident electrons.
Since particle-hole conversion changes the sign of the Lorentz force and effective mass,
the Andreev reflected holes and the electrons move coherently in the
same direction of the normal-superconducting (NS) interface. The result
is an Andreev edge state, a coherent superposition of electrons and holes,
which move in alternating skipping orbits along the NS interface \cite{Hoppe,Akhmerov}. Pseudo-magnetic
fields nevertheless reverse their direction in opposite valleys, preserving
TRS. As a consequence, when the energy of the quasiparticles, $\varepsilon$, is small compared to the 
Fermi energy, $\mu$,
reflected holes can retrace the cyclotronic path of the incident electrons, and
either form a bound state or counterpropagate along the same insulating
edge, as shown in Fig. 1a.

In this Rapid Communications, we study the transport across a NS junction
in a honeycomb lattice with a discrete spectrum of TRS LLs. The current
is carried through skipping orbits along the edge of the system both
in the normal\cite{Pouyan} and in the superconducting regions, as depicted in Fig.
1. We address the regime where the coherence length is longer than
the magnetic length. In the limit where the energy of the quasiparticles is
small compared to the separation of the LLs, we show that particle-hole
conversion produces a peak in each longitudinal conductance plateau.
Those peaks are centered around partial filling factors $\nu=4n$,
$n\in\mathbb{Z}$, when the normal LLs are half-filled, and their
height is quantized at $(n+\frac{1}{2})8e^{2}/h$. 

The Andreev edge
states carry a finite charge current per valley along the NS interface.
The valley current becomes asymptotically small at large energy ($\varepsilon\gg\mu$),
when electrons and holes move with the same group velocity. In the
opposite regime, $\varepsilon\ll\mu$, electrons and holes have opposite
group velocities along the interface, and the valley current is
finite. We find transport and spectroscopy signatures that uniquely identify proximity
induced superconductivity in TRS LLs \cite{Uchoa,Covaci}. 

\begin{figure}[b]
\begin{centering}
\includegraphics[scale=0.41]{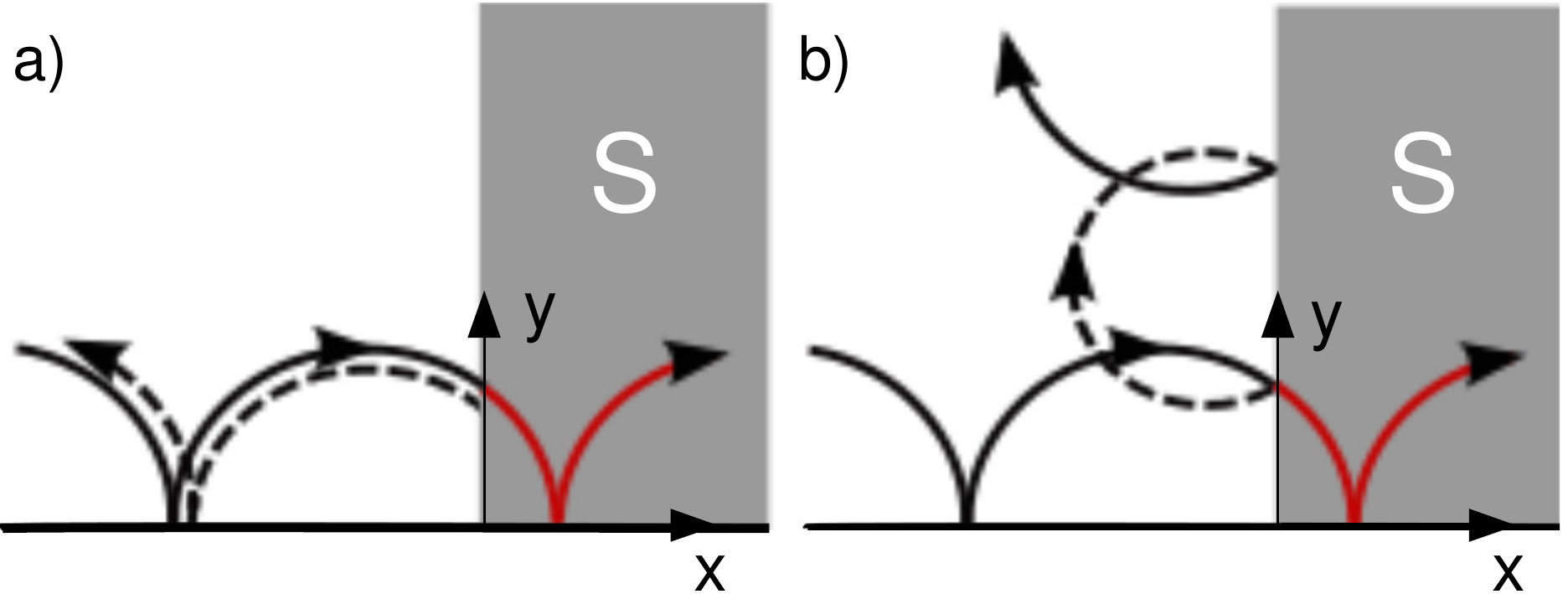}
\vspace{-0.2cm}
\par\end{centering}

\begin{centering}
\protect\caption{Semi-classical picture of the edge states in TRS LLs at the interface
with a superconducting region. Solid black lines: incident electrons
with skipping orbits. Red lines: propagating electron pairs at the
edge of the superconductor; dashed lines: Andreev reflected holes.
a) $\varepsilon\ll\mu$ regime: holes are retroreflected and retrace
the path of the incident electrons, preserving their guiding center.
They can form a bound state or counterpropagate along the same edge.
b) $\varepsilon>\mu$ regime: Andreev  holes are specularly
reflected and propagate along the NS interface as a superposition
of electrons and holes. }

\par\end{centering}

\centering{}\label{Fig:one}
\end{figure}

\emph{Hamiltonian. }In the continuum, the electronic Hamiltonian in
the presence of a pseudo magnetic-field is\cite{Antonio} 
\begin{equation}
\mathcal{H}_{0}(\mathbf{A})=\left(\begin{array}{cc}
\mathcal{H}_{+} & 0\\
0 & \mathcal{H}_{-}
\end{array}\right)=\sum_{\alpha=\pm}\nu_{\alpha}\otimes\mathcal{H}_{\alpha},\label{eq:Ho-1}
\end{equation}
where $\nu_{\alpha}=\frac{1}{2}(\nu_{0}+\alpha\nu_{z})$ is the projection
operator into the valleys $\alpha=\pm$, and 
\begin{equation}
\mathcal{H}_{\alpha}(\mathbf{A})=v(-i\nabla+\alpha\mathbf{A})\cdot\vec{\sigma}_{\alpha}-\mu\label{eq:halpha}
\end{equation}
is the Dirac Hamiltonian in each valley. $\vec{\sigma}_{\alpha}=(\sigma_{x},\alpha\sigma_{y})$
is a vector of Pauli matrices, $v$ is the Fermi velocity and $\mu$
is the chemical potential. 

Near the NS interface, the Bogoliubov-deGennes (BdG) Hamiltonian is
\begin{equation}
\mathcal{H}_{\textrm{BdG}}=\begin{pmatrix}\mathcal{H}_{0}(\mathbf{A}) & \hat{\Delta}(\mathbf{r})\\
\hat{\Delta}^{*}(\mathbf{r}) & -\mathcal{T}\mathcal{H}_{0}(\mathbf{A})\mathcal{T}^{-1}
\end{pmatrix}\label{BDG1}
\end{equation}
where \cite{Beenakker2} 
\begin{equation}
\mathcal{T}=\left(\begin{array}{cc}
0 & \sigma_{z}\\
\sigma_{z} & 0
\end{array}\right)\mathcal{C}=\nu_{x}\otimes\sigma_{z}\mathcal{C}\label{eq:time}
\end{equation}
is the time reversal symmetry operator, and $\mathcal{C}$ the charge
conjugation. Because the field $\mathbf{A}$ preserves TRS, $\mathcal{T}\mathcal{H}_{0}(\mathbf{A})\mathcal{T}^{-1}=\mathcal{H}_{0}(\mathbf{A})$.
In a singlet state, the electrons pair symmetrically across the valleys.
The simplest Ansatz for the off-diagonal term can be written as \cite{Uchoa}
\begin{equation}
\hat{\Delta}(\mathbf{r})=\Delta(\mathbf{r})\left(\begin{array}{cc}
0 & \sigma_{0}\\
\sigma_{0} & 0
\end{array}\right)=\Delta(\mathbf{r})\nu_{x}\otimes\sigma_{0}.\label{eq:gap}
\end{equation}
Assuming a sharp NS interface, the superconductor gap varies abruptly
at $x=0$, 
\begin{equation}
\Delta(\mathbf{r})=\begin{cases}
\begin{array}{c}
\Delta\\
0
\end{array} & \begin{array}{c}
\mbox{, for \ensuremath{x>0}}\\
,\mbox{ for }x<0,
\end{array}\end{cases}\label{eq:delta}
\end{equation}
which separates the normal ($x<0$) from the superconducting region
$(x>0$).

In the Landau gauge, $\mathbf{A}=(-By,0)$, the Hamiltonian in the
valleys can be written as 
\begin{equation}
\mathcal{H}_{\alpha}=\frac{v}{\sqrt{2}\ell_{B}}\alpha\left(i\partial_{\xi}\sigma_{y}+\xi\sigma_{x}\right)-\mu\label{eq:Halpha2}
\end{equation}
where $\xi=\ell_{B}k-y/\ell_{B}$ is a dimensionless coordinate with
guiding center $X_{\alpha}=\ell_{B}k$, and $\ell_{B}=\sqrt{\hbar/Be}$
is the magnetic length (restoring $\hbar$). By convention, we define
the momentum along the edge for each valley as $k_{x}=\alpha k$,
with $k>0$. The electronic wavefunction in the normal region moving
towards (away from) the interface in valley $\alpha=+$ $(-)$ takes
the form $\hat{\psi}_{\alpha}(x,y)=e^{\alpha ikx}\Psi_{\alpha}(y)$,
where $\Psi_{\alpha}(y)=(\phi_{A,\alpha},\phi_{B\alpha})$ is a two
component spinor in sublattice $A$ and $B$ of the honeycomb lattice.

\emph{Edge states. }In order describe the edge states, we introduce
a mass term potential $M(y)\nu_{0}\otimes\sigma_{z}$, where $M(y)=W$
for $y<0$ and $M=0$ for $y\geq0$. In the limit $W\to\infty$, this
potential describes the edge of the system at $y=0$. The electrons
will move in skipping orbits along the $y=0$ line under the influence
of a pseudo-magnetic field. In the normal side of the NS interface,
\begin{equation}
\left[\mathcal{H}_{\alpha}+M(y)\sigma_{z}\right]\Psi_{\alpha}=\varepsilon\Psi_{\alpha},\label{eq:Eq}
\end{equation}
where $\Psi_{\alpha}(y)$ is a two component spinor in the sublattice
basis. Multiplying Eq. (\ref{eq:Eq}) on the left by the charge conjugated
form of $\mathcal{H}_{\alpha}+M(y)\sigma_{z}$, where $\mu\to-\mu$,
this equation assumes a diagonal form, 
\begin{equation}
\left[\frac{1}{2}\partial_{\xi}^{2}-\frac{1}{2}\xi^{2}-M^{2}(y)+(\varepsilon+\mu)^{2}-\frac{1}{2}\sigma_{z}\right]\Psi_{\alpha}=0.\label{eq:squared Ham}
\end{equation}
For $y\geq0$, where $M=0$, the energy spectrum of the LLs at the
edge is 
\begin{equation}
\epsilon(k_x)=s(\sqrt{2}v/\ell_{B})\sqrt{|n(k_{x})|}-\mu\label{eq:epsilonk}
\end{equation}
 where $n(k_{x})$ is a real number and $s=\mbox{sign}[n(k_x)]$. The
corresponding eigenvectors in the two valleys are of the form 
\begin{equation}
\Psi_{\alpha,n}(\xi)=\left(\begin{array}{c}
D_{n(k_{x})-1}\left(\frac{\xi}{\sqrt{2}}\right)\\
s\alpha D_{n(k_{x})}\left(\frac{\xi}{\sqrt{2}}\right)
\end{array}\right),\label{eq:Psi2}
\end{equation}
where $D_{n(k_{x})}(x)$ are parabolic cylinder functions. The determination
of $n(k_{x})$ follows from enforcing boundary conditions at the edge
and results in a discrete number of edge states shown in Fig. 2a.
For definiteness, we consider the case of a zigzag edge, where $\phi_{B,+}(k\ell_{B}/\sqrt{2})=0$,
although similar conclusions apply to any choice of boundary conditions. 

\begin{figure}
\includegraphics[scale=0.3]{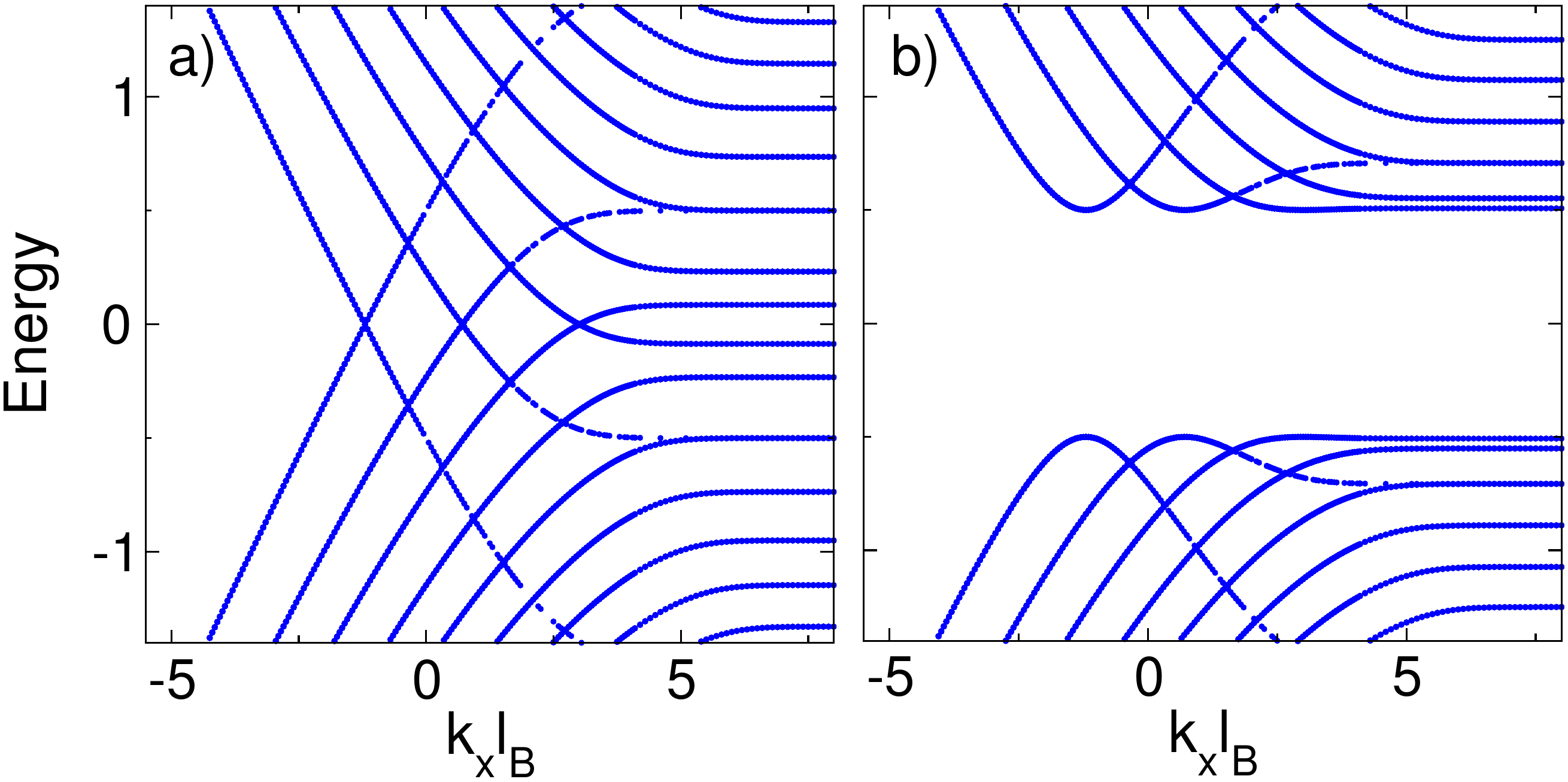}\protect\caption{Energy spectrum versus guiding center $k_{x}\ell_{B}$
at the edge ($y=0$) for $\mu=1.5$ ($\nu=4$) and $\Delta=0.5$.
Energy scales in units of $\sqrt{2}v/\ell_{B}$. a) Normal region ($x<0$).
b) superconducting region ($x>0$). }

\label{Nedge}
\end{figure}

In the superconducting region, we can decompose the BdG Hamiltonian
(\ref{BDG1}) into two identical blocks of $4\times4$ matrices, where
\begin{equation}
\begin{pmatrix}\mathcal{H}_{\alpha} & \Delta\sigma_{0}\\
\Delta^{*}\sigma_{0} & -\mathcal{H}_{\alpha}
\end{pmatrix}\Phi_{\alpha}=E\Phi_{\alpha},\label{eq:BdG2}
\end{equation}
is the reduced BdG equation and $\Phi_{\alpha}=(\Psi_{e,\alpha}\Psi_{h,\alpha})$
is a $4$ component spinor including electron and hole states. At
$y>0$, where the wavefunctions are finite ($M=0$), the solution
follows from squaring (\ref{eq:BdG2}) with the charge conjugated
BdG Hamiltonian, which results in the differential equation 
\begin{equation}
\left[\frac{1}{2}\partial_{\xi}^{2}-\frac{1}{2}\xi^{2}+\epsilon^{2}+\mu^{2}+\Delta^{2}+2\mu\mathcal{M}\right]\Phi_{\alpha}=0,\label{eq:k}
\end{equation}
where 
\[
\mathcal{M}=\left(\begin{array}{cc}
\epsilon-\frac{1}{2}\sigma_{z} & -\Delta\sigma_{0}\\
-\Delta\sigma_{0} & -\epsilon-\frac{1}{2}\sigma_{z}
\end{array}\right)
\]
is a 4$\times4$ matrix in the particle-hole basis, and $\Delta$
is assumed real. The four component spinor that satisfies Eq. (\ref{eq:k})
and hence (\ref{eq:BdG2}) is 

\begin{equation}
\Phi_{\alpha}=\left(\begin{array}{c}
\beta_{\alpha}\Psi_{\alpha,n_{s}}(\xi)\\
\beta_{-\alpha}\Psi_{\alpha,n_{s}}(\xi)
\end{array}\right)
\end{equation}
where $\beta_{\alpha}=\sqrt{\frac{1}{2}(1+\alpha\sqrt{1-\Delta^{2}/E^{2}})}$,
with $\alpha=\pm$ indexing the valleys. The energy spectrum in the superconducting edge is given
by
\begin{equation}
E(k_x)=\sqrt{\left(\bar{s}(\sqrt{2}v/\ell_{B})\sqrt{|n_{s}(k_x)|}-\mu\right)^{2}+\Delta^{2}},
\end{equation}
where $\bar{s}=\mbox{sign}[n_{s}(k_x)]$ and $n_{s}(k_x)$ is a real number
to be found from the boundary condition at the edge ($y=0$), in superconducting side.
Imposing similar boundary conditions $\phi_{B,+}(k\ell_{B}/\sqrt{2})=0$,
the energy spectrum at the superconducting edge is shown in Fig. 2b.

\emph{Transport across the NS junction.} In the normal region, the
electron and hole like excitations are decoupled. Near the NS interface,
the normal edge state can be written as a superposition of the wavefunctions
of the incident electron, $\psi_{e}^{+}(x,y)$, the reflected electron
in the opposite valley, $\psi_{e}^{-}(x,y)$, and the Andreev reflected
hole, $\psi_{h}^{-}(x,y)$. For simplicity, we assume from now on
all length scales to be in units of the magnetic length $\ell_{B}$
and all energy scales to be in units of $\sqrt{2}v/\ell_{B}$. 

The largest contribution to scattering comes from states at integer
values of $n(k)$ and $n_{s}(k)$, where the density of states is
the largest. In the four component particle-hole basis, the electron
wavefunctions are
\begin{equation}
\psi_{e,i}^{+}(x,y)=\left(\begin{array}{c}
\Psi_{+,n_{e}}\left(k_{i}-y\right)\\
\mathbf{0}
\end{array}\right)e^{ik_{i}x},
\end{equation}
and
\begin{equation}
\psi_{e,i}^{-}(x,y)=\sum_{j}^{C_{n_{e}}}r_{i,j}^{e}\left(\begin{array}{c}
\Psi_{-,n_{e}}\left(k_{j}-y\right)\\
\mathbf{0}
\end{array}\right)e^{-ik_{j}x},
\end{equation}
where the nearest integer function $n_{e}=\mbox{nint}\!\left[(\epsilon+\mu)^{2}\right]\in\mathbb{Z}$
sets the index of the highest occupied band, with $C_{n_{e}}=2n_{e}+1$
the number of channels crossing the Fermi level at the edge. The wavefunction
of the Andreev reflected hole is 
\begin{equation}
\psi_{h,i}^{-}(x,y)=\sum_{j}^{C_{n_{h}}}r_{i,j}^{A}\left(\begin{array}{c}
\mathbf{0}\\
\Psi_{-,n_{h}}\left(k_{j}-y\right)
\end{array}\right)e^{-ik_{j}x},
\end{equation}
where $n_{h}=\mbox{nint}\!\left[(\epsilon-\mu)^{2}\right]\in\mathbb{Z}$,
with $C_{n_{h}}$ equivalently the number of edge channels for hole
states.

\begin{figure}[t]
\begin{centering}
\includegraphics[scale=0.42]{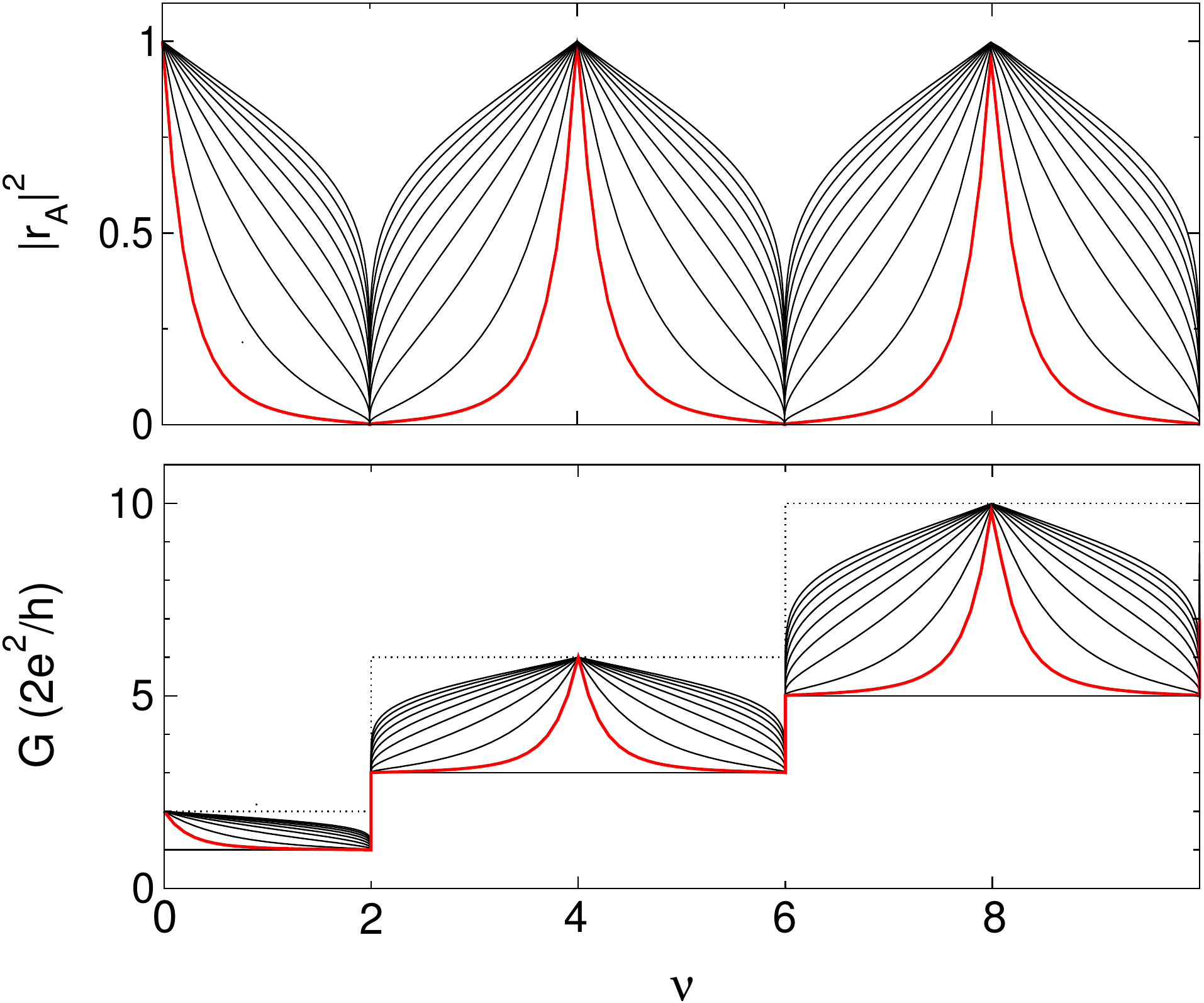} \protect\caption{Top panel: Andreev reflected hole probability $|r_{A}|^{2}$ per channel
as a function of the normal filling factor $\nu$. The superconductor
gap $\Delta$ (in units of $\sqrt{2}v/\ell_{B}$) ranges from $0.01$
(red line) to $0.2$ in steps of $0.025$. Bottom: corresponding longitudinal conductance
$G$ at the edge as a function of $\nu$. The solid line plateaus:
normal conductance ($\Delta=0$). Dotted lines: upper
bound for the conductance, which is quantized at $4e^{2}/h$. In all
curves, the temperature $T=0.02(\sqrt{2}v/\ell_{B})$.}

\par\end{centering}

\label{fig:two} 
\end{figure}

In the superconducting region, the electron-like and hole-like solutions
can be written as, 
\begin{equation}
\psi_{S,i}^{\alpha}=\sum_{j}^{C_{n_{\alpha}}}A_{i,j}^{\alpha}\left(\begin{array}{c}
\beta_{\alpha}\Psi_{\alpha,n_{+}}\left(k_{j}-y\right)\\
\beta_{-\alpha}\Psi_{\alpha,n_{+}}\left(k_{j}-y\right)
\end{array}\right)e^{\alpha ik_{j}x},
\end{equation}
where $n_{\alpha}=\mbox{nint}[\left(\sqrt{E^{2}-\Delta^{2}}+\alpha\mu\right)^{2}]\in\mathbb{Z}$.
The reflected electron and hole amplitudes can be calculated by matching
the amplitudes at the interface ($x=0$), 
\begin{equation}
\psi_{e}^{+}+\psi_{e}^{-}+\psi_{h}^{-}=\psi_{S}^{+}+\psi_{S}^{-}.\label{eq:psi}
\end{equation}

In the limit of large coherence length, $\xi=v/\Delta\gg\ell_{B}$, and $\epsilon\ll v/\ell_{B}$, we can neglect scattering
processes between different modes. Also, the number of edge channels
is the same for electrons and holes in the two sides of the junction,
$n_{e}=n_{h}=n_{\alpha}$. In this regime, the solution of Eq. (\ref{eq:psi})
is $r_{i}^{e}=0$ and $r_{i}^{A}=\beta_{+}/\beta_{-}=E/\Delta-\sqrt{E^{2}/\Delta^{2}-1}$.
When  $E<\Delta$, $r_{i}^{A}$ is complex and the
total amplitude of the Andreev reflection is $|r_{i}^{A}|^{2}=1$. 

In fig. 3, we show the amplitude of the Andreev reflection versus
the filling factor of the normal LLs. In the normal region, when the
LLs are well separated, the Fermi distribution of the quasiparticles
is equal to the filling fraction of the highest occupied LL, $(\mbox{e}^{\epsilon/T}+1)^{-1}=f(\nu)=(\nu/4+\frac{1}{2})\mbox{mod}(1)\in[0,1]$,
where $T$ is the temperature. The Andreev amplitude is $r_{i}^{A}\equiv\Theta(\nu)$,
where 
\begin{equation}
\Theta(\nu)=-\frac{|\epsilon(\nu)|}{\Delta}+\sqrt{\frac{\epsilon^{2}(\nu)}{\Delta^{2}}+1},\label{eq:Theta-1}
\end{equation}
with 
\begin{equation}
\epsilon(\nu)=T\ln\left(\frac{f(\nu)}{1-f(\nu)}\right).\label{eq:epsilon3-1}
\end{equation}

From the Blonder-Tinkham-Klapwijk formula \cite{Blonder}, the differential
conductance at the NS junction can be written as 
\begin{equation}
G=\frac{2e^{2}}{h}\sum_{i=1}^{C_{n}}(1-|r_{i}^{e}|^{2}+|r_{i}^{A}|^{2})\approx\frac{4e^{2}}{h}\!\left(\! n+\frac{1}{2}\right)\!\!\left[1+|\Theta(\nu)|^{2}\right]\!.\label{eq:G}
\end{equation}
The peaks in the differential conductance are shown in the bottom
panel of Fig. 3 as a function of the filling factor of the normal
region $\nu$. The solid line plateaus represent the conductance in
the absence of Andreev reflections. The different curves show the
conductance for different values of the normalized gap $\Delta$ ranging
from $0.01$ (red line) to 0.2 in $0.025$ steps. The peaks appear
at partial fillings $\nu=4n$, $n\in \mathbb{Z}$, and their height is quantized at
 $(2n+1)4e^{2}/h$. At those fillings, the normal LLs are particle-hole
symmetric and Andreev reflection is maximal. At integer fillings $\nu=4(n+\frac{1}{2})$,
when the Fermi level is in the middle of the LL gap, Andreev reflection
is suppressed and the conductance is quantized by half, at $(2n+1)2e^{2}/h$.

\begin{figure}
\begin{centering}
\includegraphics[scale=0.32]{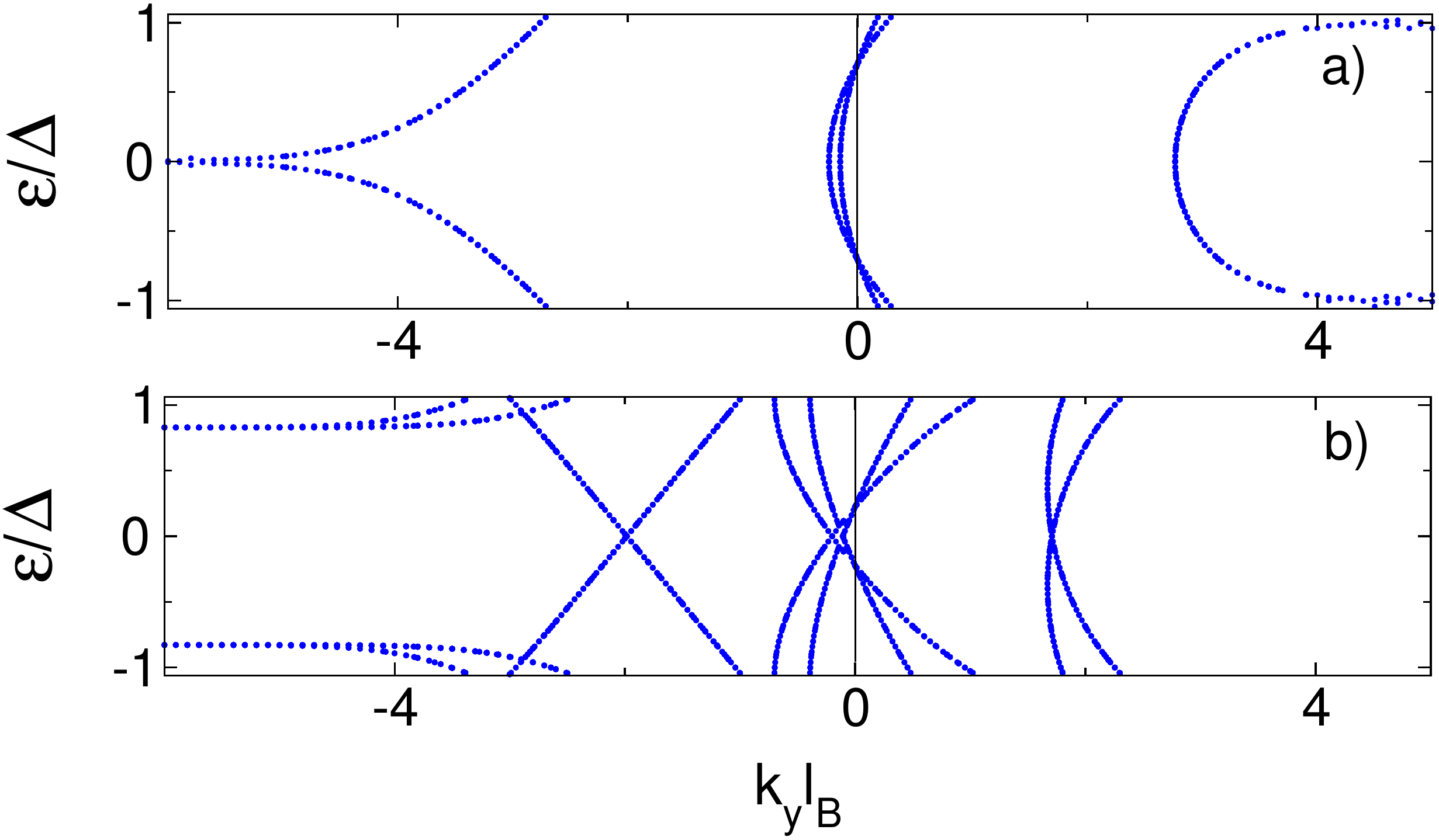}\protect\caption{Energy spectrum $\varepsilon$ along the NS interface ($x=0$)  versus guiding
center $k_{y}\ell_{B}$. Negative guiding centers ($k_{y}\ell_{B}<0$)
describe states in the normal region. $k_{y}\ell_{B}>0$ correspond
to states in the superconducting one. Left curves: normal edge states. Center right curves: low energy Andreev edge states. a) $\mu=0$ $(\nu=0$),
$\varepsilon>\mu$ regime for finite $\Delta$. Electron and hole states have the same
group velocity $v_{g}=\mbox{d}\varepsilon/\mbox{d}k_{y}$ and propagate
in the same direction of the interface; b) $\mu/\Delta=2$: electrons and Andreev reflected holes have the same
guiding center and opposite group velocities along the interface for
$\varepsilon\ll\mu$, forming an Andreev bound state.}

\par\end{centering}

\label{fig.4} 
\end{figure}

\emph{Andreev edge states.} More insight on Andreev reflection and
the electronic states near the interface can be obtained by considering
the current flowing parallel to the interface. In the configuration
where the insulating edge is zigzag, the NS interface has armchair
character. Its states are formed by a superposition of eigenstates
on both valleys. Using the Landau gauge $\mathbf{A}=(0,Bx)$, the
wavefunction in the normal side of the interface can be written as
\begin{equation}
\psi_{\parallel}(x,y)=\sum_{\alpha}\left(\begin{array}{c}
a_{e,\alpha}\Psi_{\alpha,n_{e}(\varepsilon)}\left(x-k_{y}\right)\\
a_{h,\alpha}\Psi_{\alpha,n_{h}(\varepsilon)}\left(x-k_{y}\right)
\end{array}\right)e^{\alpha ik_{y}y},\label{eq:psi_par}
\end{equation}
where $n_{e}(\varepsilon)=(\varepsilon+\mu)^{2}$ and $n_{h}(\varepsilon)=(\varepsilon-\mu)^{2}$
are real numbers. In the superconducting side, 
\begin{equation}
\psi_{S\parallel}(x,y)=\sum_{\alpha,\gamma}A_{\parallel,\gamma}^{\alpha}\left(\begin{array}{c}
\beta_{\gamma}\Psi_{\alpha,n_{\alpha}(\varepsilon)}\left(x-k_{y}\right)\\
\beta_{-\gamma}\Psi_{\alpha,n_{\alpha}(\varepsilon)}\left(x-k_{y}\right)
\end{array}\right)e^{\alpha ik_{y}y},\label{eq:psi_Spar}
\end{equation}
with $\gamma=+$$(-)$ indexing electron(hole)-like states, and $n_{\alpha}(\varepsilon)=\left(\sqrt{\varepsilon^{2}-\Delta^{2}}+\alpha\mu\right)^{2}$.
Matching the amplitudes $\psi_{\parallel}(0,y)=\psi_{S\parallel}(0,y)$
and the derivatives $\partial_{x}\psi_{\parallel}(x,y)=\partial_{x}\psi_{S\parallel}(x,y)$
at $x=0$ yields eight linear equations for eight unknown coefficients
$\mathcal{V}_{i}$, with $i=1,\ldots,8$. This set of equations can
be expressed in matrix form as $\mathcal{Q}\cdot\mathcal{V}=0$. Non-trivial
solutions require that $\mbox{Det}(\mathcal{Q})=0$. From this condition,
we numerically extract the spectrum of excitations $\varepsilon(k_{y})$
near the NS interface, as shown in Fig. 4. 

\begin{figure}[t]
\begin{centering}
\includegraphics[scale=0.35]{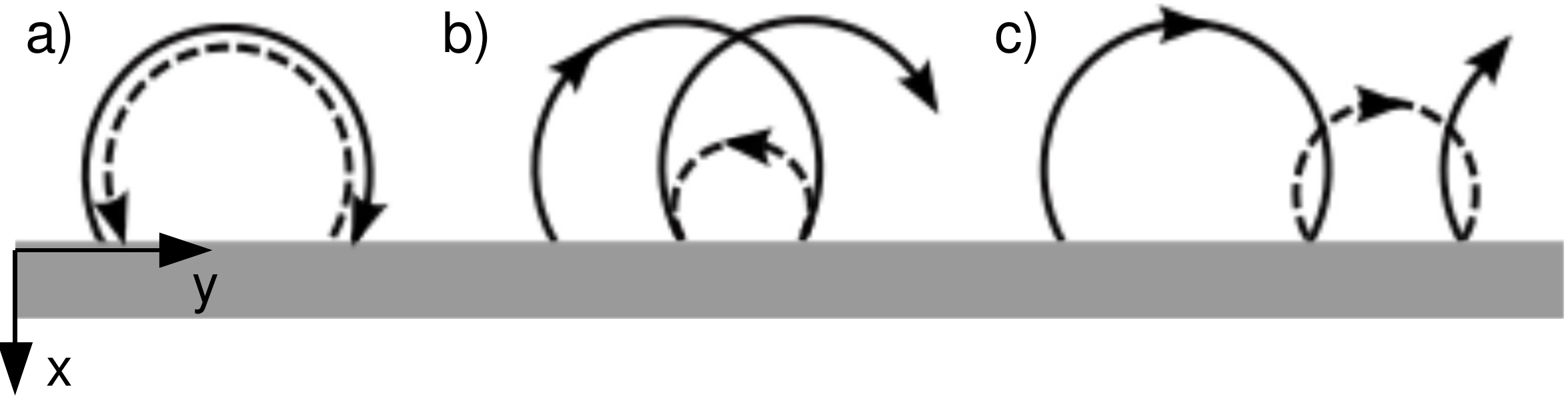}
\par\end{centering}

\begin{centering}
\vspace{-0.0cm}\protect\caption{Andreev edge states at the NS interface. Solid lines: electron cyclotronic
orbits; dashed: Andreev reflected holes. a) $\varepsilon/\mu\to0$
regime: electrons and holes form a bound state. b) Intermediate regime, $\varepsilon<\mathrm{min}(\mu,\Delta)$: electrons
are retroreflected into holes with group velocity $v_{g}=\partial\varepsilon/\partial k_{y}$
having opposite sign. c) $\varepsilon>\mu$ regime: electrons are
specularly reflected into holes, which propagate along the same direction.}

\par\end{centering}

\label{fig.5} 
\end{figure}

The interface states are an admixture of two type of modes: \emph{i)}
normal edge states formed by conventional skipping orbits moving in
one direction, and \emph{ii)} Andreev states, which are coherent superpositions
of particles and holes. The first mode is connected to bulk LL energies
in the normal region as $k_{y}\rightarrow-\infty$. The second one 
has no correspondence with the bulk LLs in the normal region,
and appears around $k_y\ell_B \sim 0$ and also inside the superconducting region, for positive guiding centers ($k_{y}>0$).

In panel 4a, we plot the interface modes at $\mu=0$ ($\nu=0$), for
finite $\Delta$. In this regime, $\varepsilon>\mu=0$, the group velocity
of the Andreev edge states $v_{g}=\partial\varepsilon/\partial k_{y}>0$
for both electrons and holes (center right curves), which are specularly reflected
at the interface \cite{Beenakker} and move in alternating skipping
orbits in the \emph{same} direction (see fig. 5c). In the limit $\varepsilon \gg \mu$ , electrons and holes have the same velocity and guiding center, and carry zero net charge per valley. In all other cases, their velocities are different, resulting in a net valley current along the NS interface. In panel 4b, we plot the energy of the modes for $\mu/\Delta=2$. In the regime $\varepsilon< \mathrm{min}(\mu,\Delta)$  we find numerically that the Andreev reflected holes and electrons have group velocities $v_g$ with opposite signs (fig. 5b). In the limit $\varepsilon/\mu \to 0$, the holes retrace the path of the incident electrons, forming an Andreev bound state schematically shown in fig. 5a. 

In summary, we have derived transport and spectroscopy signatures of proximity induced superconductivity in TRS LLs. We found the longitudinal conductance as a function of the filling factor across an NS junction, and showed that it is quantized at $(2n+1)4e^2/h$ in the $n$-th LL  at half-filling, when Andreev reflection is maximal. We also showed that the NS interface has Andreev edge states, with unique spectroscopic features.

\emph{Acknowledgements. } We thank K. Mullen and P. Carmier for discussions. BU acknowledges University of Oklahoma
and NSF Career grant DMR-1352604 for support.

\end{document}